\documentclass[preprint,aps,amsmath,amssymb,nofootinbib,tightenlines,superscriptaddress]{revtex4}

\usepackage{graphicx}
\usepackage{dcolumn}
\usepackage{bm}

\newcommand{\hc}{\ensuremath{\mathrm{h.c.}} }

\newcommand{\be}{\begin{equation}}
\newcommand{\ee}{\end{equation}}
\newcommand{\beq}{\begin{equation}}
\newcommand{\eeq}{\end{equation}}
\newcommand{\bea}{\begin{eqnarray}}
\newcommand{\eea}{\end{eqnarray}}

\newcommand{\ben}{\begin{enumerate}}
\newcommand{\een}{\end{enumerate}}

\newcommand{\bay}{\begin{array}}
\newcommand{\eay}{\end{array}}

\newcommand{\nn}{\nonumber}
\newcommand{\gev}{\ensuremath \! {\rm GeV}}
\newcommand{\Tr}{{\rm Tr}~}

\newcommand{\cvec}[2]{\left(\bay{c}#1\\#2\eay\right)}

\makeatletter       

\renewcommand{\p@subsection}{}

\renewcommand{\p@subsubsection}{}

\makeatother

\begin{document}

\vspace{3.0cm}
\preprint{\vbox
{
\hbox{WSU--HEP--1003}
}}

\vspace*{2cm}

\title{The flavor puzzle in multi-Higgs models}

\author{Andrew E. Blechman}%
\email{blechman@wayne.edu}%
\affiliation{Department of Physics and Astronomy, Wayne State University,
Detroit, MI 48201.}%
\affiliation{Michigan Center for Theoretical Physics, University of Michigan,
Ann Arbor, MI 48109.}

\author{Alexey A. Petrov}%
\email{apetrov@wayne.edu}%
\affiliation{Department of Physics and Astronomy, Wayne State University,
Detroit, MI 48201.}%
\affiliation{Michigan Center for Theoretical Physics, University of Michigan,
Ann Arbor, MI 48109.}

\author{Gagik Yeghiyan}%
\email{ye_gagik@wayne.edu}
\affiliation{Department of Physics and Astronomy, Wayne State University,
Detroit, MI 48201.}

\date{\today \\ \vspace{1in}}

\begin{abstract}
We reconsider the flavor problem in the models with two Higgs doublets.  By studying two generation toy models, we look for flavor basis independent constraints on Yukawa couplings that will give us the mass hierarchy while keeping all Yukawa couplings of the same order.  We then generalize our findings to the full three generation Standard Model.  We find that we need two constraints on the Yukawa couplings to generate the observed mass hierarchy, and a slight tuning of Yukawa couplings of order 10\%, much less than the Standard Model.  We briefly study how these constraints can be realized, and show how flavor changing currents are under control for $K-\bar{K}$ mixing in the near-decoupling limit.
\end{abstract}

\maketitle

\section{Introduction}
\renewcommand{\theequation}{1.\arabic{equation}}

The flavor problem~\cite{Georgi:1986ku} remains one of the biggest puzzles of modern particle physics.
The Standard Model (SM) of particle interactions provides a way to generate masses of quarks and
leptons, however it does not explain the apparent hierarchal structure of flavor parameters such
as fermion masses and mixing parameters~\cite{Nir:2007xn}. The ratios of the quark and lepton
masses are known experimentally, for central values given in the Particle Data Book \cite{PDG}
\bea
\frac{m_t}{m_c}\simeq 267~,\ \ &\qquad& \frac{m_c}{m_u}\simeq431~, \nn \\
\frac{m_b}{m_s}\simeq~47.5~, &\qquad& \frac{m_s}{m_d}\simeq~21~,  \label{masses} \\
\frac{m_\tau}{m_\mu}\simeq~17~,\ \ &\qquad& \frac{m_\mu}{m_e}\simeq207~.\nn
\eea
Here we use the four loop $\overline{MS}$ masses evaluated at $\mu=m_t$ for the quark masses as
defined in~\cite{mass}. In addition, the Cabibbo-Kobayashi-Maskawa (CKM) quark matrix elements
have a clear hierarchal structure, as the elements further away from the main diagonal tend to get
smaller and smaller, {\it e.g.}, $V_{ud} \sim 1$, $V_{us} \sim 0.2$, $V_{cb} \sim 0.04$, and
$V_{ub} \sim 0.004$. To add to the puzzle, the neutrino mixing matrix has a completely different
structure. In comparison, gauge couplings do not exhibit such an apparent hierarchy.

All quark and lepton masses are generated in the SM via Higgs Yukawa interactions.
For a single fermion field $\psi$ interacting with a single scalar field $\phi$,
\bea
{\cal L}_1 = - y_\psi \bar \psi_L \psi_R \phi + h.c. \to - \frac{y_\psi v}{\sqrt{2}}
\left(\bar \psi_L \psi_R + \bar \psi_R \psi_L\right),
\eea
the mass $m_\psi = y_\psi v/\sqrt{2}$ is set by the value of the Yukawa coupling, $y_\psi$, if the scalar
vacuum expectation value (vev) $v=\langle\phi\rangle$ is fixed.  This is so in the SM, where the Higgs vev
$v = 246$~GeV is fixed by the electroweak measurements, leaving a strong hierarchy in the dimensionless
Yukawa coupling sector for different quarks and leptons,
\bea
&& y_u \sim 10^{-5}, ~~y_c \sim 10^{-2},~~y_t \sim 1,
\nonumber \\
&& y_d \sim 10^{-5}, ~~y_s \sim 10^{-3},~~y_b \sim 10^{-2},
\\
&& y_e \sim 10^{-6}, ~~y_\mu \sim 10^{-3},~~y_\tau \sim 10^{-2}.
\nonumber
\eea
The reason for this hierarchy is the essence of the SM flavor problem.

One can observe that since the value of the fermion mass is given by the
product of the Higgs vev and the Yukawa coupling, the problem of the strong hierarchy
of Yukawa couplings can be made less prominent in models with several scalar fields.
For example, a hierarchy of masses of two fermions, $\psi$ and $\chi$, can be
arranged by tuning both the ratio of vevs of the scalar fields and Yukawas.
Limiting the scalar sector to two scalar fields, this can be
done in several ways. For example, each scalar can interact only with one fermion
at a time,
\bea
{\cal L}_2 = - y_\psi \bar \psi_L \psi_R \phi_1 - y_\chi \bar \chi_L \chi_R \phi_2 + \hc \label{Lagr2}
\eea
In this case, $m_\psi = y_\psi v_1/\sqrt{2}$ and $m_\chi = y_\chi v_2/\sqrt{2}$,
where $\langle \phi_1\rangle = v_1$ and $\langle \phi_2\rangle = v_2$. Here the mass
hierarchy
\beq
\frac{m_\chi}{m_\psi} = \frac{y_\chi}{y_\psi} \frac{v_2}{v_1} =  \frac{y_\chi}{y_\psi}  \tan\beta \gg 1~,
\label{hierarchy1}
\eeq
can be arranged if either $y_\chi/y_\psi \gg 1$ or $\tan\beta\equiv v_2/v_1 \gg 1$ or both.  Alternatively,
one scalar can couple to both fermions, while the other to only one,
\bea\label{Lagr2prime}
{\cal L}_2^\prime = - y_\psi \bar \psi_L \psi_R \phi_1 - y_\chi \bar \chi_L \chi_R \phi_1
- y_\chi^\prime \bar \chi_L \chi_R \phi_2 + \hc,
\eea
in which case the fermion masses are given by
\bea
m_\psi &=& y_\psi v_1/\sqrt{2}, \quad m_\chi = y_\chi v_1 /\sqrt{2}
\left(1+ \frac{y_\chi^\prime}{y_\chi}\tan\beta\right),~\mbox{and}
\nonumber \\
\frac{m_\chi}{m_\psi} &=& \frac{y_\chi}{y_\psi} \left(1+ \frac{y_\chi^\prime}{y_\chi} \tan\beta\right).
\label{hierarchy2}
\eea
Clearly, both (\ref{hierarchy1}) and (\ref{hierarchy2}) can ameliorate the fermion mass
hierarchy problem by tuning additional parameters, such as $\tan\beta$.
Models along the lines of (\ref{Lagr2}) and (\ref{Lagr2prime}) have been considered in
\cite{DK,Xu}.
However, the situation is somewhat more complicated than what one would naively expect from this simplified picture.  In general, these models are actually the same up to field redefinitions to a model with a single Higgs field getting a vacuum expectation value (vev) \cite{DH,HO}.  Therefore, if one wishes to build a model with the flavor structure leading to (\ref{hierarchy1}) or (\ref{hierarchy2}), one must supplement the above Lagrangians with additional conditions that fix which combination of Higgs fields generate a vacuum expectation value (vev).  Only after this additional constraint is specified do parameters such as $\tan\beta$ take on a physical meaning.  In models such as the minimal supersymmetric standard model (MSSM) \cite{SUSY} supersymmetry is sufficient to fix a basis for the Higgs fields; in general, however, this is an added requirement.  In this paper, we find suitable conditions by imposing constraints on the Yukawa matrices.  This fixes a special ``Higgs basis" \cite{silva1, silva2} which can be used to define $\tan\beta$.

Another complication of the SM over the above models comes from the flavor structure: while the couplings of Higgs fields to
fermions are defined in the gauge basis, the mass parameters are measured in the
mass basis. The purpose of this paper is to analyze models with an extended Higgs sector
that can be built to naturally generate the mass hierarchy.  We find basis-independent
conditions on the Yukawa matrices that ensure the hierarchy remains after rotations of
fermion basis.

We consider a class of models with two Higgs doublets,
\be
\Phi_i=\cvec{\phi_i^+}{\phi_i^0}\qquad i=1,2~.
\ee
each of which can couple to both up-type and down-type quarks and leptons. These models are
sometimes referred to as Type-III two-Higgs doublet models \cite{Cheng:1987rs,book1,III}.
The vacuum expectation values of the Higgs states can be defined as
\begin{equation}
\langle\Phi_1\rangle = \frac{1}{\sqrt{2}} \left(\begin{array}{c}
0 \\ v_1
\end{array} \right),  \hspace{0.5cm}
\langle\Phi_2\rangle = \frac{1}{\sqrt{2}} \left(\begin{array}{c}
0 \\ v_2
\end{array} \right)~. \label{m2}
\end{equation}
We assume that  $v_{1,2} > 0$ and real.  These Higgs fields then have couplings to the SM fermions
\be
-\mathcal{L}_Y=\sum_{i=1,2}\Big(\bar{Q}_L[Y_u^{(i)}] u_R\tilde{\Phi}_i + \bar{Q}_L[Y_d^{(i)}] d_R {\Phi}_i  +  \bar{L}_L[Y_\ell^{(i)}] \ell_R {\Phi}_i\Big)+\hc~. \label{Yukawas}
\ee
where $\tilde{\Phi}_i = i \sigma_2 \Phi_i^\star$ and $Y_{u,d,\ell}^{(1,2)}$
are complex generally non-Hermitian Yukawa matrices.

This paper is organized as follows. We consider two toy versions of the Standard Model
with two generations in Section~\ref{TwoGen}: first to generate the hierarchy between the first and second generation, and then the first and third generation. We then consider the realistic scenario
of all three generations in Section~\ref{ThreeGen}. Some phenomenological implications are
discussed in Section~\ref{Pheno}. Finally, we summarize our results in Section~\ref{Concl}.
The Higgs sector of the Type-III two-Higgs doublet model is reviewed in
Appendix~\ref{higgs}.  Finally, several formulae are collected in Appendix \ref{details1} and \ref{details2} for future reference.

\section{Quark mass hierarchy: two generation case}\label{TwoGen}
\renewcommand{\theequation}{2.\arabic{equation}}
\setcounter{equation}{0}

\subsection{$\tan{\beta}$ hierarchy in the 1--2 generation}

We start the quark mass hierarchy analysis by considering a toy model with two quark generations:
\begin{displaymath}
\left(\begin{array}{c}
u \\ d \end{array} \right),  \hspace{0.5cm}
\left(\begin{array}{c}
c \\ s \end{array} \right)~.
\end{displaymath}
In the most general case the Lagrangian mass terms in (\ref{Yukawas}) may be written (in the weak isospin basis) as
\begin{equation}
\left(\bar{q}_{1_L}, \bar{q}_{2_L} \right) \left[ Y^{(1)} + Y^{(2)} \tan{\beta} \right]
\left(\begin{array}{c}
q_{1_R} \\ q_{2_R} \end{array} \right) v \cos{\beta} + {\rm h.c.}~, \label{q1}
\end{equation}
where $q_1 =  u,d$; $q_2 = c,s$; $\tan{\beta} = v_2/v_1$; and we assume throughout this paper that $\tan{\beta} \gg 1$.
$Y^{(1)}$ and $Y^{(2)}$
are $2 \times 2$ complex non-Hermitian Yukawa matrices of the quark interactions with the Higgs doublets $\Phi_1$ and $\Phi_2$
respectively.
It is also convenient to define the total Yukawa matrix,
\begin{equation}
Y = Y^{(1)} + Y^{(2)} \tan{\beta}~, \label{q3}
\end{equation}
which is diagonalized by the rotation
\begin{equation}
V_L Y V_R^\dagger = \left(\begin{array}{cc}
y_1 & 0 \\ 0 & y_2 \end{array} \right)~, \label{q4}
\end{equation}
with the quark masses related to the eigenvalues as
\begin{equation}
m_{q_{1,2}} = |y_{1, 2}| v \cos{\beta}~, \label{q5}
\end{equation}
and\footnote{In the two generation case, this matrix is just the Cabibbo matrix, but the generalization to CKM is clear.} $V_{u_L} V_{d_L}^\dagger = V_{CKM}$.
Our aim is to find some $U(2)$ invariant conditions on the Yukawa
matrices that assure having a hierarchy in the eigenvalues $y_1$ and
$y_2$ and hence in the quark masses.

For $2 \times 2$ matrices the $U(2)$ invariants are related to
traces and determinants of those matrices. Rigorously speaking, only the
traces and determinants of Hermitian matrices are invariant under
$U(2)$ rotations: for instance, the traces and determinants of
$Y Y^\dagger$ and $Y^\dagger Y$. Note that
\begin{eqnarray}
V_L Y Y^\dagger V_L^\dagger = \left(\begin{array}{cc}
|y_1|^2 & 0 \\ 0 & |y_2|^2 \end{array} \right)~, \label{q8} \\
V_R Y^\dagger Y V_R^\dagger = \left(\begin{array}{cc}
|y_1|^2 & 0 \\ 0 & |y_2|^2 \end{array} \right)~. \label{q9}
\end{eqnarray}

Yet, dealing with the products $Y Y^\dagger$ and $Y^\dagger Y$ would make
our analysis too involved.  For the two generation case,
it is more instructive to generate the quark mass hierarchy, studying the matrices
$Y$, $Y^{(1)}$, $Y^{(2)}$ by themselves. We will however discuss briefly what
the conditions imposed on $Y$, $Y^{(1)}$ and/or $Y^{(2)}$ invariants imply
on $Y Y^\dagger$ and its components. This is going to be useful for the
realistic scenario with three quark (or lepton) generations.

As the matrices $Y$, $Y^{(1)}$ and $Y^{(2)}$ are non-Hermitian,
one must be careful when dealing
with the traces and determinants. Notice first that the traces of $Y$, $Y^{(1)}$
and $Y^{(2)}$ are not invariant under $U(2)$ rotations. For instance, the diagonal elements of $Y$
in the weak isospin basis are related to that in the quark mass basis by (no sum over $i$)
\begin{equation}
Y^m_{ii} = V_{L_{ij}} Y_{jk} V^\star_{R_{i k}}~. \label{q10}
\end{equation}
So $\Tr Y^m \neq \Tr  Y$ as $\sum_iV^\star_{R_{i k}} V_{L_{ij}} \neq \delta_{kj}$.

On the other hand, for the determinants we have
\begin{eqnarray}
&& \det{Y^m} = e^{i (\Phi_L - \Phi_R)} \det{Y}~, \label{q11} \\
\nonumber
&& \det{Y^{(1) m}} = e^{i (\Phi_L - \Phi_R)} \det{Y^{(1)}}~, \\
\nonumber
&& \det{Y^{(2) m}} = e^{i (\Phi_L - \Phi_R)} \det{Y^{(2)}}~,
\end{eqnarray}
where $e^{i \Phi_L} = \det{V_L}$ and $e^{i \Phi_R} = \det{V_R}$.
In other words, the determinants of $Y$, $Y^{(1)}$ and $Y^{(2)}$ are only
multiplied by some phase factor under $U(2)$ rotations. Thus the absolute
values of the determinants are rotational invariants. This allows one
to use $Y$, $Y^{(1)}$ and $Y^{(2)}$ determinants to impose some $U(2)$
rotational invariant
conditions on the Yukawa matrices and generate the desired quark mass
hierarchy.

Here we impose the condition\footnote{Up to this point, $\tan\beta$ is not a physical parameter (see the discussion in Appendix \ref{higgs}), but once we impose this constraint on the Yukawa matrices, this ambiguity is lost.}
\begin{equation}
\det{Y^{(2)}} = 0~. \label{q6}
\end{equation}
Certainly, this condition is invariant under $U(2)$ rotations. By imposing this condition, one generates the hierarchy $y_2 \sim y_1  \tan{\beta}$.
To see this, consider the eigenvalue equation for the total matrix in Equation (\ref{q3})
\begin{equation}
y^2 -   (\Tr Y) \ y + \det{Y} = 0~. \label{q7}
\end{equation}
Generally speaking, $\Tr Y$, $\det{Y}$ and hence $y_1$, $y_2$ are
complex. Yet, in the quark mass basis one redefines quark phases
so that $y_1 > 0$ and $y_2 > 0$ with both real. As $q_2$ corresponds to
heavier quark states $c$ and $s$, we will choose $y_2 > y_1$.

As
\begin{equation}
y_1 + y_2 = \Tr Y = \Tr[Y^{(1)} + Y^{(2)} \tan{\beta}]
\sim O(Y^{(2)} \tan{\beta})~,  \label{q12}
\end{equation}
one infers that
\begin{equation}
y_2 \sim O(Y^{(2)} \tan{\beta}) \label{q13}~.
\end{equation}
On the other hand
\begin{equation}
y_1 y_2 = \det{Y} = \det{Y^{(1)}} +
\varepsilon_{i j} \varepsilon_{k l} \left( Y_{i k}^{(1)} Y_{j l}^{(2)}
+ Y_{i k}^{(2)} Y_{j l}^{(1)} \right) \tan{\beta} +
\det{Y^{(2)}} \tan^2 {\beta}~. \label{q14}
\end{equation}
Condition (\ref{q6}) on the $Y^{(2)}$ determinant assures that $O(\tan^2{\beta})$ terms on
the r.h.s. of (\ref{q14}) vanish. Thus,
\begin{equation}
y_1 y_2 \sim O(Y^{(1)} Y^{(2)} \tan{\beta})~. \label{g15}
\end{equation}
Hence, combining (\ref{q13})~and~(\ref{g15}) one gets
\begin{equation}
y_1 \sim O(Y^{(1)})~, \label{q16}
\end{equation}
where $O(Y^{(1)})$ denotes the order of the $Y^{(1)}$ matrix elements -- during
our analysis we assume that this matrix elements are of the same order (at least the
diagonal ones).
Thus, as it follows from (\ref{q13})~and~(\ref{q16}),
\begin{equation}
y_2 \sim y_1 \tan{\beta}~, \label{g17}
\end{equation}
provided that there is no hierarchy in the elements of the matrices
$Y^{(1)}$ and $Y^{(2)}$.

The exact solutions of the eigenvalue equation~(\ref{q7}) may be written as
\begin{eqnarray}
\nonumber
y_{1,2} = \frac{1}{2} \Biggl\{ Y^{(1)}_{11} + Y^{(1)}_{22} +
\left(Y^{(2)}_{11} + Y^{(2)}_{22} \right) \tan{\beta} \mp
\Biggl[\left(Y^{(1)}_{11} + Y^{(1)}_{22} +
\left(Y^{(2)}_{11} + Y^{(2)}_{22} \right) \tan{\beta} \right)^2 \\
- 4 \left(Y^{(1)}_{11} Y^{(1)}_{22} - Y^{(1)}_{12} Y^{(1)}_{21} \right) -
4 \left(Y^{(1)}_{11} Y^{(2)}_{22} + Y^{(2)}_{11} Y^{(1)}_{22} - Y^{(1)}_{12} Y^{(2)}_{21}
- Y^{(2)}_{12} Y^{(1)}_{21}\right)  \tan\beta\Biggr]^{1/2} \Biggr\}~. \label{q18}
\end{eqnarray}
Expanding (\ref{q18}) in terms of $1/\tan{\beta}$ power series, one gets
\begin{eqnarray}
y_1&\approx&\frac{Y^{(1)}_{11} Y^{(2)}_{22} + Y^{(2)}_{11} Y^{(1)}_{22} - Y^{(1)}_{12} Y^{(2)}_{21}
- Y^{(2)}_{12} Y^{(1)}_{21}}{Y^{(2)}_{11} + Y^{(2)}_{22}}~, \label{q19} \\
&& \nonumber \\
\nonumber
y_2&\approx&\left( Y^{(2)}_{11} + Y^{(2)}_{22} \right) \tan{\beta}  + Y^{(1)}_{11} + Y^{(1)}_{22} \\
&-&\frac{Y^{(1)}_{11} Y^{(2)}_{22} + Y^{(2)}_{11} Y^{(1)}_{22} - Y^{(1)}_{12} Y^{(2)}_{21}
- Y^{(2)}_{12} Y^{(1)}_{21}}{Y^{(2)}_{11} + Y^{(2)}_{22}}~. \label{q20}
\end{eqnarray}
The $O(\tan{\beta})$ hierarchy in the values of $y_1$ and $y_2$ is apparent. Also, in
terms of the mass ratios one gets
\begin{equation}
\frac{m_{q_2}}{m_{q_1}} \approx \frac{\left| Y^{(2)}_{11} + Y^{(2)}_{22} \right|^2 \tan{\beta}}
{|Y^{(1)}_{11} Y^{(2)}_{22} + Y^{(2)}_{11} Y^{(1)}_{22} - Y^{(1)}_{12} Y^{(2)}_{21}
- Y^{(2)}_{12} Y^{(1)}_{21}|}~. \label{q21}
\end{equation}

Note that $O(\tan{\beta})$ hierarchy alone is insufficient to reproduce quark mass ratios for
the both types of
quarks (as well as charged leptons). Recall that for the central values of the
fermions masses one has
\begin{displaymath}
\frac{m_s(m_t)}{m_d(m_t)} \simeq 21, \hspace{0.5cm}
\frac{m_c(m_t)}{m_u(m_t)} \simeq 431, \hspace{0.5cm}
\frac{m_\mu}{m_e} \simeq 207~.
\end{displaymath}
Choosing e.g. $\tan{\beta} = 20$, one can reproduce the strange to down quark mass ratio.
Yet, to reproduce the other ratios, an additional reduction of the denominator in
(\ref{q21}) is necessary, by imposing some conditions on the relevant Yukawa couplings.
The simplest way to do it is to assume that  $(Y_u^{(1)})_{ij} \sim 0.05 \Tr Y_u^{(2)}$ and
$(Y_\ell^{(1)})_{ij} \sim 0.1 \Tr Y_\ell^{(2)}$.
There is nothing technically unnatural in
imposing such conditions, and this small tuning is drastically reduced from the usual SM Yukawas. Moreover, as it follows from our analysis, we have an expansion
in terms of $\frac{Y^{(1)}}{Y^{(2)} \tan{\beta}}$ rather than of $1/\tan{\beta}$. In what
follows, these
assumptions on the up-quark and charged lepton Yukawa matrices do not spoil our derivations.

Thus, imposing the rotationally invariant condition (\ref{q6})
on the $Y^{(2)}$ determinant, one is able to reproduce the
first and second generation quark and lepton mass ratios, without assuming a large family
hierarchy in the couplings with the Higgs doublets.

To see what the imposed condition on the $Y^{(2)}$ determinant implies on the quark
interactions with the Higgs doublets, note that in addition to the mass and weak isospin bases, two
additional quark bases exist that are relevant:
\begin{itemize}
\item{basis~(a) where the matrix $Y^{(1)}$ is diagonal; this basis is related to the
weak isospin basis as
\begin{equation}
\left(\begin{array}{c} q_1^{(a)} \\ q_2^{(a)} \end{array} \right)_{L,R}  =
V_{L,R}^{(a)} \left(\begin{array}{c} q_1 \\ q_2 \end{array} \right)_{L,R}~,
\hspace{1.5cm}
V_{L}^{(a)} Y^{(1)} V_{R}^{(a) \dagger} \equiv Y^{(1) a} = \left(
\begin{array}{cc} y_1^{(1)} & 0 \\ 0 & y_2^{(1)} \end{array} \right)~.
\label{q22}
\end{equation}}
\item{basis~(b) where the matrix $Y^{(2)}$ is diagonal; this basis is related to the
weak isospin basis as
\begin{equation}
\left(\begin{array}{c} q_1^{(b)} \\ q_2^{(b)} \end{array} \right)_{L,R} =
V_{L,R}^{(b)} \left(\begin{array}{c} q_1 \\ q_2 \end{array} \right)_{L,R}~,
\hspace{1.5cm}
V_{L}^{(b)} Y^{(2)} V_{R}^{(b) \dagger} \equiv Y^{(2) b} = \left(
\begin{array}{cc} y_1^{(2)} & 0 \\ 0 & y_2^{(2)} \end{array} \right)~.
\label{q23}
\end{equation}}
\end{itemize}

As the condition is imposed on $Y^{(2)}$ determinant, it is natural to consider the quark
interactions with the Higgs doublets in basis~(b). In that basis, condition~(\ref{q6})
implies
\begin{equation}
Y^{(2) b} = \left(\begin{array}{cc} 0 & 0 \\
0 & y_2^{(2)} \end{array} \right)   \hspace{0.5cm} \text{or}
\hspace{0.5cm}
Y^{b} = \left(\begin{array}{cc} Y^{(1) b}_{11} &
Y^{(1) b}_{12} \\ Y^{(1) b}_{21} & ~~Y^{(1) b}_{22} +
y_{2}^{(2)} \tan{\beta}
\end{array} \right)~. \label{q24}
\end{equation}
In other words, in basis~(b) the second Higgs doublet interacts with the second generation quarks
only. The first generation quarks interact with each other and with the second generation quarks
solely due to exchange of $\Phi_1$. This interaction scheme is depicted below.
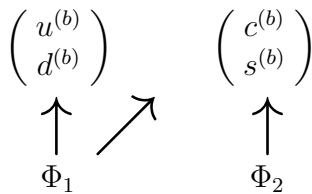
\begin{eqnarray*}
\left(\begin{array}{c} u^{(b)} \\ d^{(b)} \end{array} \right) \hspace{1.1cm} &
\left(\begin{array}{c} c^{(b)} \\ s^{(b)} \end{array} \right) \\
\text{\Huge $\uparrow$ $\nearrow$}  \hspace{0.5cm} & \text{\Huge $\uparrow$}  \label{q25} \\
\Phi_1 \hspace{1.6cm} & \Phi_2
\end{eqnarray*}

This scheme is very similar in spirit to ``texture" models in \cite{textures,textures2,textures3}.  The big difference between these models and ours is that they assume this structure in the gauge basis, whereas we impose the basis independent condition (\ref{q6}) and {\it derive} this scenario.  However, as we see below, basis (b) is generally distinct from the gauge basis, and this will have important consequences in what follows.

It is also worth mentioning that in terms of the Yukawa matrix elements in basis~(b), the formula for the quark mass ratios looks like
\begin{equation}
\frac{m_{q_2}}{m_{q_1}} \approx \frac{ |Y^{(2) b}_{22}| \tan{\beta} }
{ |Y^{(1) b}_{11}| } = \frac{ |y^{(2)}_{2}| \tan{\beta} }
{ |Y^{(1) b}_{11}| }~. \label{q26}
\end{equation}
A similar interaction scheme and formula for the mass ratio may be derived in basis~(b) for the charged
lepton families as well.

One may choose basis~(b) to coincide with the weak isospin basis, by assuming that $V_{d_L}^{(b)} =
V_{d_R}^{(b)} = V_{u_L}^{(b)} = V_{u_R}^{(b)} = V^{(b)}$ and redefining the isospin basis as
\begin{displaymath}
\left(\begin{array}{c} d \\ s \end{array} \right) \to V^{(b)}
\left(\begin{array}{c} d \\ s \end{array} \right)~, \hspace{1.5cm}
\left(\begin{array}{c} u \\ c \end{array} \right) \to V^{(b)}
\left(\begin{array}{c} u \\ c \end{array} \right)~.
\end{displaymath}
However, such a scenario does not seem to be realistic. It is not hard to infer from
(\ref{q24})~and~(\ref{q26}) that basis~(b) is transformed to the quark mass basis by
means of rotation angles $\sim m_{q_1}/m_{q_2} \ll \theta_C$, where $\theta_C$ is the
Cabibbo angle with $\sin{\theta_C} \approx 0.2259$. Thus, generating the Cabibbo mixing
properly within a scenario with coinciding weak isospin basis and basis~(b) is very unlikely.
One should rather have the weak isospin basis distinctly different from basis~(b) and
with $\Phi_2$ interacting (in the isospin basis) with both
the first and second quark generations, however
with the Yukawa couplings being constrained by condition
(\ref{q6}).

On the other hand, basis~(b) differs only slightly from the quark mass basis: as discussed,
these two bases are related by small rotations ($\sim m_d/m_s \sim 0.05$ and
$\sim m_u/m_c \sim 0.002$ for the down and up sectors respectively; also if extending
our analysis to the charged lepton sector, $\sim m_e/m_\mu \sim 0.005$). Thus, the
interaction scheme within basis~(b) presented above in (\ref{q25}),
is {\it nearly true} in the mass basis as well. Namely, one has
$Y^{(2)m}_{11}, Y^{(2)m}_{12}, Y^{(2)m}_{21} \sim \left(m_{q_1}/m_{q_2} \right)
Y^{(2)m}_{22}$ and
$Y^{(2)m}_{11}, Y^{(2)m}_{12}, Y^{(2)m}_{21} \sim Y^{(1)m}_{11}/\tan{\beta},
Y^{(1)m}_{22}/\tan{\beta}$, since we assumed $Y^{(1)m}_{11}, Y^{(1)m}_{22} \sim
\left(m_{q_1} \tan{\beta}/m_{q_2}\right) Y^{(2)m}_{22}$, as discussed above. In other words,
within the quark mass basis, the interaction of $\Phi_2$ with the first generation quarks
is greatly suppressed as compared both to that of $\Phi_2$ with the second generation
quarks and to that of the other doublet, $\Phi_1$, with both generations of quarks.

Thus, we conclude that imposing the rotationally invariant
condition~(\ref{q6}) on the $Y^{(2)}$ matrix determinant for
$\tan{\beta} \gg 1$ gives the desired quark mass hierarchy, as well as an interaction
scheme where, within the quark mass basis, the Higgs doublet $\Phi_2$ interacts
predominantly with the second generation quarks, while the other Higgs doublet $\Phi_1$
interacts equally with both quark generations. Extending this picture for the charged lepton
generations is also straightforward.

To conclude this subsection, we discuss what condition
(\ref{q6}) implies when considering the Hermitian product $(Y Y^\dagger)$; we will need this
when switching to the three-generation case as well as in the next subsection.
Note that in addition to the constraints $\det{(Y^{(2)} Y^{(2)\dagger})} = \det{(Y^{(2)} Y^{(1)\dagger})} =
\det{(Y^{(1)} Y^{(2)\dagger})} = 0$,
condition (\ref{q6}) also implies
\begin{eqnarray}
\nonumber
\det{\left[\left(Y^{(1)} Y^{(2)\dagger} + Y^{(2)} Y^{(1)\dagger} \right) \tan{\beta}
+ Y^{(2)} Y^{(2)\dagger} \tan^2{\beta} \right]}  \\
= \det{\left[Y^{(1)} Y^{(2)\dagger} +
Y^{(2)} Y^{(1)\dagger}\right]} \tan^2{\beta}~,  \label{q29}
\end{eqnarray}
which is easily proven in basis~(b).
The product $Y Y^\dagger$ may be presented as
\begin{equation}
Y Y^\dagger = Y^{(1)} Y^{(1)\dagger} + Y^{(1)} Y^{(2)\dagger} \tan{\beta}  +
Y^{(2)} Y^{(1)\dagger} \tan{\beta}
+ Y^{(2)} Y^{(2)\dagger} \tan^2 {\beta}~. \label{q27}
\end{equation}
Generally, for large $\tan\beta$, $\det{(Y Y^\dagger)} \sim O(\tan^4{\beta})$, however as condition (\ref{q29}) is imposed, one gets
\begin{equation}
\det{(Y Y^\dagger)} \sim O(\tan^2{\beta})~. \label{q28}
\end{equation}

\subsection{$\tan^2{\beta}$ hierarchy in the 1--3 generation}

Having just one scheme for generating the fermion mass hierarchy is insufficient to reproduce
all three quark and charged lepton masses. In order to reproduce properly the first
and second and the first and third family mass ratios, at least two mechanisms for
generating the mass hierarchy are needed. The first mechanism has been discussed in the previous
subsection. The natural candidate for the second mechanism is the one that generates an
$O(\tan^2{\beta})$ hierarchy. Indeed, the quark mass ratios may be presented as:
\begin{eqnarray}
&& A\equiv\frac{m_s(m_t)}{m_d(m_t)} \simeq 21~, \hspace{1.2cm}
\frac{m_b(m_t)}{m_d(m_t)} \simeq 2.26 \times A^2~,  \label{q30} \\
&& B\equiv\frac{m_c(m_t)}{m_u(m_t)} \simeq 431~, \hspace{1cm}
\frac{m_t(m_t)}{m_u(m_t)} \simeq 0.62 \times B^2~.  \label{q31}
\end{eqnarray}
Thus, the third to first generation mass ratios may be presented as the second to first generation
mass ratios squared multiplied by some $\mathcal{O}(1)$ factors. These factors may easily be
generated by appropriately choosing the values of the Yukawa matrix elements without imposing
any family hierarchy on the Yukawa couplings.

In this subsection we continue to study the toy model with two quark generations, however
we now look for a $U(2)$ invariant condition that generates an $O(\tan^2{\beta})$
hierarchy in the total Yukawa matrix eigenvalues and hence in the quark masses.
Subsequently, $q_2$ now denotes $t$ or $b$ quark states.

An $O(\tan^2 \beta)$ hierarchy in the quark masses may be generated by imposing the rotationally
invariant condition
\begin{equation}
|\det{Y}| = |\det{Y^{(1)}} |~.  \label{q32}
\end{equation}
This condition assures that
\begin{equation}
y_1 y_2 = \det{Y} \sim O\left( (Y^{(1)})^2 \right)~, \label{q33}
\end{equation}
which, combined with $y_2 \sim O(Y^{(2)} \tan{\beta})$ as shown in Equation (\ref{q13}),
yields
\begin{equation}
y_1 \sim O\left( \frac{\left(Y^{(1)}\right)^2}{Y^{(2)} \tan{\beta}} \right)~, \label{q34}
\end{equation}
and subsequently,
\begin{equation}
\frac{y_2}{y_1} \sim \tan^2{\beta} \label{q35}~.
\end{equation}
The exact solutions of the eigenvalue equation (\ref{q7}) is now
\begin{eqnarray}
\nonumber
y_{1,2}&=&\frac{1}{2} \Biggl\{ Y^{(1)}_{11} + Y^{(1)}_{22} +
\left(Y^{(2)}_{11} + Y^{(2)}_{22} \right) \tan{\beta} \\
&\mp&\Biggl[\left(Y^{(1)}_{11} + Y^{(1)}_{22} +
\left(Y^{(2)}_{11} + Y^{(2)}_{22} \right) \tan{\beta} \right)^2
- 4 \det{Y} \Biggr]^{1/2} \Biggr\}~, \label{q36}
\end{eqnarray}
which, after expansion in powers of $1/\tan{\beta}$,
may be rewritten as
\begin{eqnarray}
y_1&\approx&\frac{\det{Y}}{\left(Y^{(2)}_{11} + Y^{(2)}_{22}\right) \tan{\beta}}+\mathcal{O}(\tan^{-2}\beta)~, \label{q37} \\
&& \nonumber \\
y_2&\approx&\left( Y^{(2)}_{11} + Y^{(2)}_{22} \right) \tan{\beta}  + Y^{(1)}_{11} + Y^{(1)}_{22}
-\frac{\det{Y}}{\left(Y^{(2)}_{11} + Y^{(2)}_{22}\right)\tan{\beta}}+\mathcal{O}(\tan^{-2}\beta)~. \label{q38}
\end{eqnarray}
In general, there is an ambiguity in solutions (\ref{q37}) and (\ref{q38}) because of an
unknown phase in
\begin{displaymath}
\det{Y} = e^{i \phi} \det{Y^{(1)}} = e^{i \Phi} \left(Y^{(1)}_{11} Y^{(1)}_{22} -
Y^{(1)}_{12} Y^{(1)}_{21} \right)~.
\end{displaymath}
Yet, in the mass basis where $y_1 > 0$, $y_2 > 0$ and hence
$\det{Y} = |\det{Y^{(1)}}| > 0$, this ambiguity is removed. More generally,
for large $\tan\beta$, the last term in the expression for $y_2$ may
be neglected, and for $y_1$ this problem is avoided by considering the
absolute values of the eigenvalues, as only the absolute values have physical
meaning. Then
\begin{eqnarray}
|y_1|&\approx&\frac{\left|Y^{(1)}_{11} Y^{(1)}_{22} -
Y^{(1)}_{12} Y^{(1)}_{21} \right|}{\left|Y^{(2)}_{11} + Y^{(2)}_{22}\right| \tan{\beta}}~, \label{q39} \\
&& \nonumber \\
|y_2|&\approx&\left| Y^{(2)}_{11} + Y^{(2)}_{22} \right| \tan{\beta}~. \label{q40}
\end{eqnarray}
Subsequently,
\begin{equation}
\frac{m_{q_2}}{m_{q_1}} \approx \frac{\left| Y^{(2)}_{11} + Y^{(2)}_{22} \right|^2
\tan^2{\beta}}{\left|Y^{(1)}_{11} Y^{(1)}_{22} -
Y^{(1)}_{12} Y^{(1)}_{21} \right|}~. \label{q41}
\end{equation}
Thus, imposing condition (\ref{q32}) on $|\det{Y}|$, one gets the desired
$O(\tan^2{\beta})$ hierarchy in the total Yukawa matrix eigenvalues and
subsequently on the quark mass ratios.

To see what this condition on $|\det{Y}|$ implies on the quark
interactions with the Higgs doublets, it is convenient to
rewrite (\ref{q32}) in the following form:
\begin{equation}
\det{(Y Y^\dagger)} = \det{\left(Y^{(1)} Y^{(1) \dagger}\right)}~. \label{q42}
\end{equation}
Comparing to (\ref{q27}), $\tan{\beta}$ dependent terms in the expression
for $\det{(Y Y^\dagger)}$ must vanish to satisfy condition (\ref{q42}).
In general, this may occur in different ways. Yet, for $\tan{\beta} \gg 1$,
the natural way to satisfy (\ref{q42}) is to demand for
the $\tan{\beta}$-dependent terms to vanish to all orders in $\tan{\beta}$.

It has already been discussed in the previous subsection that the vanishing of
$O(\tan^4{\beta})$ and $O(\tan^3{\beta})$ terms in $\det{(Y Y^\dagger)}$
may be assured by imposing condition (\ref{q6}) on $\det{Y^{(2)}}$.
This means that we have again the interaction scheme where $\Phi_2$ interacts
with the heaviest family of quarks -- exactly in basis~(b) and predominantly in
the mass basis.

Yet, as condition (\ref{q32}) or equivalently (\ref{q42}) on $\det{Y}$
is much stronger than (\ref{q6}), one may expect that the interaction
scheme corresponding to $O(\tan^2{\beta})$ quark mass hierarchy is more
constrained than that discussed in the previous subsection.
To see this, one may rewrite the Hermitian product $Y Y^\dagger$ in
basis~(b) in the following form (provided that $\det{Y^{(2)}} = 0$):
{\footnotesize
\be\begin{split}
&Y^b Y^{b \dagger} = \\ &
\left( \begin{array}{c c} |Y^{(1) b}_{11}|^2 + |Y^{(1) b}_{12}|^2, &
Y^{(1) b \star}_{21} Y^{(1) b}_{11} + Y^{(1) b \star}_{22} Y^{(1) b}_{12} +
y^{(2) \star}_2 Y^{(1) b}_{12} \tan{\beta} \\
Y^{(1) b}_{21} Y^{(1) b \star}_{11} + Y^{(1) b}_{22} Y^{(1) b \star}_{12} +
y^{(2)}_2 Y^{(1) b \star}_{12} \tan{\beta}, &
|Y^{(1) b}_{21}|^2 + |Y^{(1) b}_{22}|^2 +
2 Re\left[y^{(2)}_2 Y^{(1) b \star}_{22} \right] \tan{\beta} +
|y_2^{(2)}|^2 \tan^2{\beta} \end{array} \right)
\end{split}\ee}

The conditions for $O(\tan^2{\beta})$ and $O(\tan{\beta})$
terms in $\det{(Y Y^\dagger)}$ to vanish in the rotational invariant form are
respectively
(provided that $\det{Y^{(2)}} = 0$)
\begin{eqnarray}
\nonumber
\det{\left(Y^{(1)} Y^{(2) \dagger} + Y^{(2)} Y^{(1) \dagger} \right)} \tan^2{\beta}
+ \det{\left(Y^{(1)} Y^{(1) \dagger} + Y^{(2)} Y^{(2) \dagger} \tan^2{\beta} \right)} \\
- \det{\left(Y^{(1)} Y^{(1) \dagger}\right)} = 0~, \label{q44} \\
\nonumber
\det{\left[Y^{(1)} Y^{(1) \dagger} +
\left(Y^{(1)} Y^{(2) \dagger} + Y^{(2)} Y^{(1) \dagger} \right) \tan{\beta}
\right]} \hspace{3cm} \\
- \det{\left(Y^{(1)} Y^{(2) \dagger} + Y^{(2)} Y^{(1) \dagger} \right)} \tan^2{\beta}
- \det{\left(Y^{(1)} Y^{(1) \dagger}\right)} = 0~.  \label{q45}
\end{eqnarray}
It is a matter of algebra to show that these two conditions in basis~(b) become
\begin{equation}
Y^{(1) b}_{11} = 0~. \label{q43}
\end{equation}
In other words, the rotationally invariant condition (\ref{q32}) not only leads to an
$O(\tan^2{\beta})$ hierarchy in the quark (and charged lepton) masses, but also implies that in
basis~(b) the lightest generation quarks do not interact with the doublet $\Phi_2$ and
interact with the doublet $\Phi_1$ only via transitions to the heavier
generation quarks.
This scheme is also nearly true in the quark mass basis, since as before, basis~(b) differs from the mass basis by small rotation angles
($\sim m_d/m_b \sim 0.001$; $\sim m_u/m_t \sim 10^{-5}$; $\sim m_e/m_\mu \sim 0.0005$).

\section{Quark mass hierarchy: three generation case}\label{ThreeGen}
\renewcommand{\theequation}{3.\arabic{equation}}
\setcounter{equation}{0}

\subsection{Conditions on Yukawa Matrices}

Having the mass hierarchy generation mechanisms at hand, we may now turn to the realistic
three generation model.  For the three generation case, the mass terms in the Lagrangian may be written as
\begin{equation}
\left(\bar{q}_{1_L}, \bar{q}_{2_L}, \bar{q}_{3_L} \right) \left[ Y^{(1)} + Y^{(2)} \tan{\beta} \right]
\left(\begin{array}{c}
q_{1_R} \\ q_{2_R} \\ q_{3_R} \end{array} \right) v \cos{\beta} + {\rm h.c.}~, \label{q46}
\end{equation}
where $Y^{(1)}$ and $Y^{(2)}$ are now $3 \times 3$ complex generally non-Hermitian matrices.
The total Yukawa matrix is still given by (\ref{q3}), and
\begin{equation}
V_L Y V_R^\dagger = \left(\begin{array}{ccc}
y_1 & 0 & 0 \\ 0 & y_2 & 0 \\
0 & 0 & y_3 \end{array} \right)~, \label{q48}
\end{equation}
with the quark masses related to the eigenvalues as
\begin{equation}
m_{q_{i}} = |y_{i}| v \cos{\beta}, \hspace{0.5cm}  i = 1, 2, 3~. \label{q49}
\end{equation}
The eigenvalue equation is now
\begin{equation}
y^3 - (\Tr Y) \ y^2 + ({\det}_2 Y) \ y - \det{Y} = 0~, \label{q50}
\end{equation}
where
\begin{equation}
{\det}_2{Y} = \sum_{i < j} \left(Y_{ii} Y_{jj} - Y_{ij} Y_{ji} \right)~, \label{q51}
\end{equation}
is the sum of all the second order diagonal minors of $Y$.
In the mass basis, one may choose real $y_1 > 0$, $y_2 > 0$ and
$y_3 > 0$, by redefining the quark phases.
As $q_1 = u, d$; $q_2 = c, s$; $q_3 = t, b$; we assume $y_3 >
y_2 > y_1$.

If no condition is imposed on the Yukawa matrices, one gets
\begin{eqnarray*}
&& y_1 + y_2 + y_3 = \Tr Y=
\Tr\left(Y^{(1)} + Y^{(2)} \tan{\beta}\right) \sim O(Y^{(2)} \tan{\beta})~, \\
&& y_1 y_2 + y_1 y_3 + y_2 y_3 =
{\det}_2{Y} = {\det}_2\left(Y^{(1)} + Y^{(2)} \tan{\beta}\right) \sim
O\left((Y^{(2)})^2 \tan^2{\beta} \right)~,  \\
&& y_1 y_2 y_3 = \det{Y} = \det\left(Y^{(1)} + Y^{(2)} \tan{\beta}\right) \sim
O\left((Y^{(2)})^3 \tan^3{\beta} \right)~,
\end{eqnarray*}
and subsequently
\begin{displaymath}
y_3 \sim y_2 \sim y_1 \sim O(Y^{(2)} \tan{\beta})~.
\end{displaymath}
Yet our aim is to find $U(3)$ invariant constraints on the matrix elements that yield
\begin{eqnarray}
{\det}_2{Y} = {\det}_2\left(Y^{(1)} + Y^{(2)} \tan{\beta}\right) \sim
O\left(Y^{(1)} Y^{(2)} \tan{\beta} \right)~, \label{q52} \\
\det{Y} = \det\left(Y^{(1)} + Y^{(2)} \tan{\beta}\right) \sim
O\left((Y^{(1)})^3 \right)~, \label{q53}
\end{eqnarray}
and thus
\begin{eqnarray}
&& y_3 \sim O(Y^{(2)} \tan{\beta})~, \label{q54} \\
&& y_2 \sim O(Y^{(1)})~, \label{q55} \\
&& y_1 \sim O\left(\frac{Y^{(1)}}{Y^{(2)} \tan{\beta}}\right)~. \label{q56}
\end{eqnarray}

The relevant condition on $\det{Y}$ is still given by (\ref{q32}) or,
equivalently, by (\ref{q42}).  However, there is a problem with imposing conditions on
${\det}_2{Y}$, ${\det}_2{Y^{(1)}}$ or ${\det}_2{Y^{(2)}}$: these quantities are not
invariant under $U(3)$ rotations. Thus, at this point we cannot use matrices $Y$,
$Y^{(1)}$ and $Y^{(2)}$ anymore. Rather we have to proceed to the Hermitian
product $Y Y^\dagger$ (or $Y^\dagger Y$) and its components.

For $Y Y^\dagger$ we have
 \begin{equation}
 V_L Y Y^\dagger V_L^\dagger = \left(\begin{array}{ccc}
|y_1|^2 & 0 & 0 \\ 0 & |y_2|^2 & 0 \\
0 & 0 & |y_3|^2
 \end{array} \right)~, \label{q57}
\end{equation}
and the eigenvalue equation is now
\begin{equation}
|y|^6 - (\Tr\left( Y Y^\dagger \right)) \ |y|^4 + ({\det}_2\left( Y Y^\dagger\right )) \
|y|^2 - \det\left( Y Y^\dagger\right ) = 0~, \label{q58}
\end{equation}
and thus
\begin{eqnarray}
\nonumber
|y_1|^2 + |y_2|^2 + |y_3|^2 = \Tr\left( Y Y^\dagger \right) =
\Tr\biggl[ Y^{(1)} Y^{(1) \dagger} + \left( Y^{(1)} Y^{(2) \dagger} +
Y^{(2)} Y^{(1) \dagger} \right) \tan{\beta} \\
+ Y^{(2)} Y^{(2) \dagger} \tan^2{\beta}  \biggr]
\sim O\left(|Y^{(2)}|^2 \tan^2{\beta} \right)~, \label{q59}
\end{eqnarray}
and (with the use of condition (\ref{q42}))
\begin{equation}
|y_1|^2 |y_2|^2 |y_3|^2 = \det\left( Y Y^\dagger \right) =
\det\left( Y^{(1)} Y^{(1) \dagger} \right) \sim O\left(|Y^{(1)}|^6 \right)~. \label{q60}
\end{equation}
Note that for $3 \times 3$ Hermitian matrices the sum of the second
order diagonal minors {\it is} invariant under $U(3)$ rotations and therefore may be
used to derive the missing condition that leads to the desired hierarchy of the
eigenvalues. This condition is
\begin{equation}
{\det}_2 \left(Y^{(2)} Y^{(2) \dagger} \right) = 0~. \label{q61}
\end{equation}
Apart from the fact that this condition implies
$\det \left(Y^{(2)} Y^{(2) \dagger} \right) = 0$, one also gets
\begin{equation}
|y_1|^2 |y_2|^2 + |y_1|^2 |y_3|^2 +
|y_2|^2 |y_3|^2 = {\det}_2 \left(Y Y^{\dagger} \right)
\sim O\left(|Y^{(1)}|^2 |Y^{(2)}|^2 \tan^2{\beta} \right)~. \label{q62}
\end{equation}
As before, one can show this working in basis~(b), where the matrix
$Y^{(2)}$ is diagonal. With condition (\ref{q61}), one has
\begin{equation}
 Y^{(2) b} Y^{(2) \dagger b} = \left(\begin{array}{ccc}
 0 & 0 & 0 \\
 0 & 0 & 0 \\
 0 & 0 & |y_3^{(2)}|^2 \end{array} \right) \hspace{0.5cm}  \Rightarrow
 \hspace{0.5cm}
 Y^{(2) b} = \left(\begin{array}{ccc}
 0 & 0 & 0 \\
 0 & 0 & 0 \\
 0 & 0 & y_3^{(2)} \end{array} \right)~. \label{q63}
\end{equation}
Because of the importance for our analysis, we also present explicitly
the total Yukawa matrix $Y$ and $Y Y^\dagger$ in basis~(b) in Appendix \ref{details1}.
With the use of (\ref{q63}) and the formulae in the Appendix, proving that
${\det}_2 \left(Y Y^{\dagger} \right)
\sim O\left(|Y^{(1)}|^2 |Y^{(2)}|^2 \tan^2{\beta} \right)$ is
straightforward.

One infers from Eqs.~(\ref{q59}), (\ref{q60}), (\ref{q62}), for $|y_3|^2 >
|y_2|^2 > |y_1|^2$,
\begin{eqnarray}
&& |y_3|^2 \sim O\left(|Y^{(2)}|^2 \tan^2{\beta} \right)~, \label{q66} \\
&& |y_2|^2 |y_3|^2 \sim
O\left(|Y^{(1)}|^2 |Y^{(2)}|^2 \tan^2{\beta} \right)~, \label{q67} \\
&& |y_1|^2 |y_2|^2 |y_3|^2 \sim
O\left(|Y^{(1)}|^6 \right)~, \label{(q68}
\end{eqnarray}
or
\begin{eqnarray}
&& |y_3|^2 \sim O\left(|Y^{(2)}|^2 \tan^2{\beta} \right)~, \label{q69} \\
&& |y_2|^2 \sim
O\left(|Y^{(1)}|^2  \right)~, \label{q70} \\
&& |y_1|^2  \sim
O\left(\frac{|Y^{(1)}|^4}{|Y^{(2)}|^2 \tan^2 {\beta}} \right)~. \label{q71}
\end{eqnarray}
This is the desired hierarchy in the values of $y_1$,
$y_2$ and $y_3$.

Formulae (\ref{q69})-(\ref{q71}) determine only the order of magnitude of
$|y_1|$, $|y_2|$ and $|y_3|$ qualitatively. Finding the
most general solution of the cubic eigenvalue equation (\ref{q58})
is not easy. However, if $|y_3|^2 \gg |y_2|^2 \gg
|y_1|^2$, as it follows from Eqs.~(\ref{q69})-(\ref{q71}), one
gets
\begin{eqnarray}
&& |y_3|^2 \approx \Tr\left( Y Y^\dagger \right)~, \label{q72} \\
&& |y_2|^2 \approx \frac{{\det}_2\left( Y Y^\dagger \right)}
{\Tr\left( Y Y^\dagger \right)}~, \label{q73} \\
&& |y_1|^2 \approx \frac{\det\left( Y Y^\dagger \right)}
{{\det}_2\left( Y Y^\dagger \right)}~, \label{q74}
\end{eqnarray}
where $\det\left( Y Y^\dagger \right)$ is given by (\ref{q42}) and,
following the accuracy of the approach, one should leave only the leading-order in
$\tan{\beta}$ terms in the expressions for $\Tr\left( Y Y^\dagger \right)$ and
${\det}_2\left( Y Y^\dagger \right)$.  The resulting formulae for the $|y_i|^2$ and the subsequent mass ratios are given in Appendix \ref{details2}.

For $\tan{\beta} = 20$, the down-type quark mass ratios
\begin{equation}
\frac{m_s(m_t)}{m_d(m_t)} \simeq 1.05 \tan{\beta}, \hspace{0.5cm}
\frac{m_b(m_t)}{m_s(m_t)} \simeq 2.38 \tan{\beta}, \hspace{0.5cm}
\frac{m_b(m_t)}{m_d(m_t)} \simeq 2.5 \tan^2{\beta}, \label{q81}
\end{equation}
may be reproduced by choosing the elements of matrices $Y_d^{(1)}$
and $Y_d^{(2)}$ to be of the same order while satisfying
the imposed rotationally invariant conditions. Numerically, the
elements of these matrices must be chosen appropriately to
reproduce the finite factors in front of $\tan{\beta}$ and
$\tan^2{\beta}$ in (\ref{q81}), however no family hierarchy in
the down-quark Yukawa interactions is needed.

To reproduce the up-type quark mass ratios,
\begin{equation}
\frac{m_c(m_t)}{m_u(m_t)} \simeq 21.6 \tan{\beta}, \hspace{0.5cm}
\frac{m_t(m_t)}{m_c(m_t)} \simeq 13.4 \tan{\beta}, \hspace{0.5cm}
\frac{m_t(m_t)}{m_u(m_t)} \simeq 290 \tan^2{\beta}, \label{q82}
\end{equation}
some weak tuning must be imposed on the denominators of
(\ref{q78})-(\ref{q80}). Like in the toy models with two generations, the easiest way
to do this is to assume
\begin{displaymath}
|(Y_u^{(1)})_{i j}|^2 \sim 0.01 ~\Tr\left( Y_u^{(2)} Y_u^{(2) \dagger} \right)~.
\end{displaymath}
As discussed, this condition does not spoil our derivations: in
fact our expansion is in powers of $\frac{Y^{(1)}}{Y^{(2)} \tan{\beta}}$
rather than in powers of $1/\tan{\beta}$. Again, no large family hierarchy in the
Yukawa interactions is needed.

Thus, imposing condition (\ref{q42}) on the $Y Y^\dagger$ determinant and
condition (\ref{q61}) on the sum of the $Y^{(2)} Y^{(2) \dagger}$
second order diagonal minors, one is able to reproduce the actual
ratios of the quark masses, without imposing a large family hierarchy
on the Yukawa interactions of the quarks with the Higgs doublets.

While no family hierarchy in the quark Yukawa interactions is
assumed in our model, the imposed rotational invariant conditions
(\ref{q42}) and (\ref{q61}) certainly have an impact on
interactions, as discussed in the previous section. As before,
it is convenient to examine this impact in basis~(b) where the
matrix $Y^{(2)}$ is diagonal. In this basis, as it follows from
Eq.~(\ref{q63}), only the third generation quarks interact with
$\Phi_2$, as depicted in the scheme below.
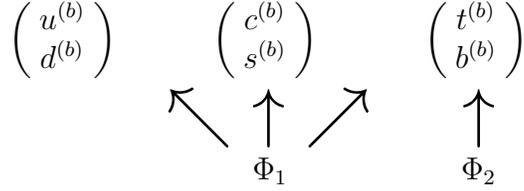
\begin{eqnarray*}
\left(\begin{array}{c}
u^{(b)} \\ d^{(b)}  \end{array} \right) \hspace{1.2cm}
\left(\begin{array}{c}
c^{(b)} \\ s^{(b)}  \end{array} \right) \hspace{1.2cm} &
\left(\begin{array}{c}
t^{(b)} \\ b^{(b)}  \end{array} \right) \\
\text{\Huge $\nwarrow$ $\uparrow$ $\nearrow$}  \hspace{0.5cm} & \text{\Huge $\uparrow$}  \label{q83} \\
\Phi_1 \hspace{1.6cm} & \Phi_2
\end{eqnarray*}
This interaction scheme remains nearly true in the mass basis too, as
\begin{equation}
q_3^{(b)} \approx q_3^{(m)}~, \label{q84}
\end{equation}
with accuracy of $O(m_{q_2}/m_{q_3}) \sim O\left(\frac{Y^{(1)}}{Y^{(2)} \tan{\beta}}\right)$ terms.
This stems from the fact that
\begin{eqnarray}
\nonumber
(Y^{b} Y^{b \dagger})_{33} \approx |y_3^{(2)}|^2 \tan^2{\beta} \gg
(Y^{b} Y^{b \dagger})_{13}, (Y^{b} Y^{b \dagger})_{23} \sim O\left(Y^{(1)}
y_3^{(2)} \tan{\beta}\right) \\
\gg (Y^{b} Y^{b \dagger})_{11},
(Y^{b} Y^{b \dagger})_{12}, (Y^{b} Y^{b \dagger})_{22} \sim O\left((Y^{(1)})^2\right)~.
\label{q85}
\end{eqnarray}

So far the analysis has been conducted along the same lines as within the previous section
for the toy two generation models. Yet, as the three generation case is more involved in general,
it is natural to expect that some differences in the analysis still may occur. One of them
is related to the constraints on the light quark interactions with $\Phi_1$, due to
condition (\ref{q42}) on $\det\left(Y Y^\dagger\right)$. For the two-generation case
condition (\ref{q42}) gives (\ref{q43}) or equivalently that the lightest generation
quarks interact in basis~(b) with $\Phi_1$ only via transitions to the heaviest generation
quarks; this remains nearly true in the mass basis as well. For the three generation
case
condition (\ref{q42}) places constraints on combinations of the Yukawa
couplings rather than on only one of them. For instance, one gets
\begin{equation}
Y^{(1) b}_{11} Y^{(1) b}_{22} - Y^{(1) b}_{12} Y^{(1) b}_{21} = 0~. \label{q86}
\end{equation}
The scenario where for example $Y^{(1) b}_{11} = Y^{(1) b}_{12} = 0$,
{\it i.e.:} the first generation quarks in basis~(b) interact
with $\Phi_1$ only via transitions to heavier generation quarks, is only one particular
scenario that satisfies (\ref{q86}).
More generally, (\ref{q86}) may
be satisfied in any scenario with
$Y^{(1)b}_{ij}$ tuned appropriately.

Most importantly, any condition expressed in terms of $Y^{(1)}$ matrix elements
in basis~(b) changes drastically when rotating to the mass basis. This
is because unlike the toy models of the previous section, in the three-generation case {\it basis~(b)
and the mass basis are not related by small rotations} as far as the
first two generation mixing angles are concerned. In other words, if neglecting
the third generation mixing with two others, one has
\begin{eqnarray}
&& q_1^{(m)} \approx q_1^{(b)} \cos{\theta_{12}^{(b \to m)}}  +
q_2^{(b)} \sin{\theta_{12}^{(b \to m)}}~, \label{q87} \\
&& q_2^{(m)} \approx - q_1^{(b)} \sin{\theta_{12}^{(b \to m)}}  +
q_2^{(b)} \cos{\theta_{12}^{(b \to m)}}~, \label{q88}
\end{eqnarray}
where $\theta_{12}^{(b \to m)}$ is not small in general. This stems from the fact that the
elements of the $2 \times 2$ upper sub-matrix of the  matrix $Y^b Y^{b \dagger}$ are in
general the same order, as it follows from formula (\ref{q65}) in the appendix. Thus,
$\theta_{12}^{(b \to m)}$ should not be small in general for the hierarchy in the
values of $m_{q_2}$  and $m_{q_1}$ to be generated.

One may in principle have $\theta_{12}^{(b \to m)} \sim \theta_C$ if one assumes
a slight hierarchy,
$Y^{(1) b}_{11} \sim 0.25 Y^{(1) b}_{22}$.
The advantage of allowing such a
hierarchy is that unlike the two generation toy models,
basis~(b) may naturally coincide with the weak isospin basis; the
necessary conditions for this to occur have been discussed in the previous section.
In that case, the interaction scheme depicted above (\ref{q84}) is valid both in the mass basis
and in the isospin basis.

In summary, when imposing the rotationally invariant condition (\ref{q42}) on the $Y Y^\dagger$ determinant and
(\ref{q61}) on the sum of the $Y^{(2)} Y^{(2) \dagger}$
second order diagonal minors, in addition to reproducing the actual
ratios of the quark masses, one derives a quark-to-Higgs interacting scheme where in basis~(b) the Higgs doublet $\Phi_2$
interacts only with the third generation of quarks. This scheme remains nearly true in
the mass basis as well. Also, if one allows a slight hierarchy in the elements of
the upper $2 \times 2$ sub-matrix of the matrix $Y^{(1)}$, one may choose basis~(b)
to coincide with the weak isospin basis. In that case the derived interaction scheme
is the one both within the isospin basis (precisely) and within the mass basis
(approximately). Notice also that the imposed rotationally invariant conditions
imply some conditions on (rather complicated) combinations of the $Y^{(1)}$
matrix elements.

We complete this section by considering the charged lepton mass problem. One may proceed in the same way as for the quarks. For $\tan{\beta} = 20$,
\begin{equation}
\frac{m_\mu}{m_e} \simeq 10.4 \tan{\beta}, \hspace{0.5cm}
\frac{m_\tau}{m_\mu} \simeq 0.85 \tan{\beta}, \hspace{0.5cm}
\frac{m_\tau}{m_e} \simeq 8.8 \tan^2{\beta}. \label{q95}
\end{equation}
The $\mathcal{O}(1)$ coefficient in front of $\tan{\beta}$ for the ratio $\frac{m_\tau}{m_\mu}$
indicates that the elements of the matrices $Y_\ell^{(1)}$ and $Y_\ell^{(2)}$
must be of the same order, as one can infer from Eq.~(\ref{q79}) .
Yet, to reproduce the coefficient 10.4
in front of $\tan{\beta}$ for the ratio $\frac{m_\mu}{m_e}$,
the elements of the matrix $Y_\ell^{(1)}$ must be tuned appropriately for
$\det\left(Y_{\ell}^{(1)} Y_{\ell}^{(1) \dagger} \right)$ to be suppressed, as it
follows from (\ref{q78}).

\subsection{More on Basis (b)}\label{BasisB}

Because of its crucial importance, basis~(b) and its physical
meaning, as well as the meaning of condition (\ref{q61}), deserve more detailed discussion. If one assumes for the Higgs masses
$m_{A^0}, m_{H^+}, m_{H^0} \gg m_{h^0}$, so that flavor changing neutral currents (FCNCs) are suppressed,
then for the CP-even Higgs rotation angles defined in Appendix \ref{higgs}, one has $\alpha \approx \beta - \pi/2$.
If $\tan{\beta} \gg 1$, (\ref{m6}) and (\ref{m7}) (ignoring Goldstone modes) may be approximated by
\begin{equation}
\Phi_1 \approx \left( \begin{array}{c}
 - H^+  \\
\frac{1}{\sqrt{2}} \left[v_1  + H^0 - i A^0
\right] \end{array} \right)~, \label{q91}
\end{equation}
\begin{equation}
\Phi_2 \approx \left( \begin{array}{c}
 0 \\
\frac{1}{\sqrt{2}} \left[v  + h^0  \right] \end{array} \right)~.
\label{q92}
\end{equation}
To this approximation, $\Phi_2$ is the SM Higgs doublet, while $\Phi_1$ is new physics (NP). Thus, basis~(b)
is the basis where the SM Yukawa matrix is diagonal.

In our model, the family symmetry is broken in two steps.
Quark interactions with the SM Higgs doublet
$\Phi_2$ break $U(3)$ quark family symmetry down to $U(2)$. If
only $\Phi_2$ gets a vev, then only the top and bottom quarks would acquire masses,
while other quarks would remain massless. Yet
interactions of the NP Higgs doublet $\Phi_1$ with quarks break the family symmetry completely and
generate both the first two generation quark masses and the CKM mixing. Thus, in the scenario considered here, the up, down,
strange and charm quark interactions with the Higgs particles as well as the CKM mixing are
predominantly beyond the Standard Model physics. Yet, the Yukawa interactions of the first two
generation quarks with the Higgs doublets are still suppressed, due to the NP Higgs masses being at TeV or even higher scales.

This interpretation  of the  model assumes that the weak isospin basis coincides with basis~(b).
On the other hand, if this model is an effective theory originating from a more fundamental
theory at TeV or higher scales, then the weak isospin basis may be different from basis~(b). Note that
our results based on the rotationally invariant conditions are independent of how these two bases
are related to each other.

There are strong reasons to believe that the two-Higgs doublet model discussed here is an
effective theory that originates from
a more fundamental theory that occurs at TeV or higher scales. For instance,
having the NP Higgs masses at TeV or higher scales requires the mass parameters $\mu_1$,
$\mu_2$ and $\mu_3$ of the Higgs potential to have magnitudes of the order of TeV or higher
scales as well.  A possible explanation of the scale of these parameters may be the existence of a
gauge singlet scalar field $S$, with interactions
\begin{equation}
\mathcal{L}_S\supset\lambda_1^S |S|^2 \left(\Phi_1 \Phi_1^\dagger \right) + \lambda_2^S |S|^2
\left(\Phi_2 \Phi_2^\dagger \right) +
\left(\lambda_3^S S^2 \left(\Phi_1 \Phi_2^\dagger \right) + {\rm h.c.} \right)~, \label{q93}
\end{equation}
with
\begin{equation}
\mu_1^2 = \lambda_1^S \langle S \rangle^2, \hspace{0.5cm}
\mu_2^2 = \lambda_2^S \langle S \rangle^2, \hspace{0.5cm}
\mu_3^2 = \lambda_3^S \langle S \rangle^2~, \label{q94}
\end{equation}
and $\langle S \rangle \gg v = 246~\gev$.

Another reason to believe there is a more fundamental theory at higher scales is that presently
we are able to clearly interpret only condition (\ref{q61}) on the sum of the $Y^{(2)} Y^{(2) \dagger}$
second order diagonal minors through the importance of basis (b). The meaning of the other condition, (\ref{q42}) on the
$Y Y^\dagger$ determinant and the resulting constraints on the $Y^{(1)}$ matrix elements
remain obscure.

\section{Phenomenological Implications: Flavor-Changing Processes and $K-\bar{K}$ Mixing}\label{Pheno}
\renewcommand{\theequation}{4.\arabic{equation}}
\setcounter{equation}{0}

Let us now consider flavor changing processes.  As mentioned in Appendix \ref{higgs}, in the limit that $m_A\gg v$, these are naturally suppressed, but we would like to see this explicitly.
To do that, we write out the Yukawa interactions in a very suggestive way:

\bea
-\mathcal{L}_Y&=&\bar{Q}_L [Y_u] u_R\tilde{\Phi}_1 + \bar{Q}_L [Y_u^{(2)}] u_R\tilde{\Psi}  \nn \\
& &+~\bar{Q}_L [Y_d] d_R {\Phi}_1  + \bar{Q}_L [Y_d^{(2)}] d_R{\Psi} +{\rm h.c.}~.
\eea
where $\tilde{\Phi}_1 = i \sigma_2 \Phi_1^\star$, $\tilde{\Psi} = i \sigma_2 \Psi^\star$ and $Y_{u,d}$ are the total Yukawa matrices for the up-type and down-type quarks, defined in (\ref{q3}).
We have also defined the linear combination of Higgs fields
\be
\Psi=\Phi_2-\Phi_1\tan\beta~,
\ee
and we are only considering the physical Higgs fields ((\ref{m6}) and (\ref{m7}) minus the vevs).  It should be clear that this is the same as our original Yukawa interactions, but the first term in each line is proportional to the mass matrices and is therefore flavor diagonal in the mass basis by construction.  Therefore all the tree level flavor-changing processes in the Higgs sector couple to the $\Psi-$combination of Higgs fields and appear in the second term on each line.  Also note that all FCNCs are coming from $Y^{(2)}$, whose off diagonal elements in the mass basis are naturally small due to (\ref{q61}).  Notice that this is consistent with the interpretation of Section \ref{BasisB}.

With FCNCs at tree level, we can apply constraints from various flavor standard candles, such as meson mixing and electric dipole measurements~\cite{Isidori:2010kg}.  Since we have already shown that we can suppress FCNCs in various regions of parameter
space, we will only consider $K-\bar{K}$ mixing in this paper (which is typically the strongest constraint), and leave the other flavor observables for future research \cite{paper2}.

To study $K-\bar{K}$ mixing, we consider the effective Lagrangian
\begin{displaymath}
\mathcal{L}_{\rm eff}=\frac{1}{m_{h^0}^2}\sum_i C_i\mathcal{O}_i~;
\end{displaymath}
we will use the operator basis of \cite{BMZ}, where they define the four-quark operators ($i,j$ are color indices):
\begin{align}
\mathcal{O}_1&=(\bar{d}^i_L\gamma^\mu s^i_L)(\bar{d}^j_L\gamma_\mu s^j_L)~,&\widetilde{\mathcal{O}}_1&=(\bar{d}^i_R\gamma^\mu s^i_R)(\bar{d}^j_R\gamma_\mu s^j_R)~,\nn \\
\mathcal{O}_2&=(\bar{d}^i_Rs^i_L)(\bar{d}^j_Rs^j_L)~,&\mathcal{O}_3&=(\bar{d}^i_Rs^j_L)(\bar{d}^j_Rs^i_L)~,\nn  \\
\widetilde{\mathcal{O}}_2&=(\bar{d}^i_Ls^i_R)(\bar{d}^j_Ls^j_R)~,&\widetilde{\mathcal{O}}_3&=(\bar{d}^i_Ls^j_R)(\bar{d}^j_Ls^i_R)~,\nn  \\
\mathcal{O}_4&=(\bar{d}^i_Rs^i_L)(\bar{d}^j_Ls^j_R)~,&\mathcal{O}_5&=(\bar{d}^i_Rs^j_L)(\bar{d}^j_Ls^i_R)~.
\end{align}
There are also dipole operators, but these are irrelevant at tree level.  For $K-\bar{K}$ mixing there are three Higgs exchange diagrams at tree level that give
\bea
\mathcal{M}_1&=&\frac{i}{2} (Y^{(2)*}_{d21})^2  \langle \Psi^{0 \star} \Psi^{0 \star} \rangle\langle K^0|\mathcal{O}_2|\bar{K}^0\rangle~, \\
\mathcal{M}_2&=&\frac{i}{2} (Y^{(2)}_{d12})^2\langle  \Psi^{0} \Psi^{0} \rangle\langle K^0|\widetilde{\mathcal{O}}_2|\bar{K}^0\rangle~, \\
\mathcal{M}_3&=&i \ (Y^{(2)}_{d12}Y^{(2)*}_{d21}) \langle \Psi^0 \Psi^{0 \star} \rangle\langle K^0|\mathcal{O}_4|\bar{K}^0\rangle ~,
\eea
where the $\Psi^0$ propagators are for the neutral Higgs states (that is, the lower component of the doublet).  It is a straightforward exercise to expand out the Higgs propagators using the mass basis defined in Appendix \ref{higgs} and this allows us to write down the tree level Higgs contributions to the matching conditions at the Higgs mass scale\footnote{For this paper we will chose $\mu_h=m_{h^0}$ and ignore the errors of order $\log\left(\frac{m_{\rm heavy}}{m_{\rm light}}\right)$, but for the sake of generality we keep $\mu_h$ arbitrary in these formulae.} $\mu_h$:
\small
\bea
C_2(x,y;\mu_h)&=&-\frac{1}{4}(Y^{(2)*}_{d21})^2\left[(\cos\alpha+\sin\alpha\tan\beta)^2+\frac{(\sin\alpha-\cos\alpha\tan\beta)^2}{x}-\frac{\sec^2\beta}{y}\right]~,\label{c2}\\
C_4(x,y;\mu_h)&=&-\frac{1}{2}(Y^{(2)}_{d12}Y^{(2)*}_{d21})\left[(\cos\alpha+\sin\alpha\tan\beta)^2+\frac{(\sin\alpha-\cos\alpha\tan\beta)^2}{x}+\frac{\sec^2\beta}{y}\right]~,\label{c4}
\eea
\normalsize
where $x\equiv m_{H^0}^2/m_{h^0}^2$ and $y\equiv m_{A^0}^2/m_{h^0}^2$; $\widetilde{C}_2$ is the same as $C_2$ with $Y^{(2)*}_{d21}\rightarrow Y^{(2)}_{d12}$ and
\be
C_{1}(\mu_h)=\widetilde{C}_1(\mu_h)=C_{3}(\mu_h)=\widetilde{C}_3(\mu_h)=C_5(\mu_h)=0~.
\ee
Notice that in the limit $m_{A^0}\rightarrow\infty$, the heavy Higgs contributions vanish\footnote{Recall the Heavy CP-even Higgs field mass also grows with $m_{A^0}$ from (\ref{m9}).}.  Furthermore, in the same limit, $\alpha\simeq\beta-\pi/2$ and a little trigonometry shows that the light Higgs contribution also vanishes.  Therefore, there are no contributions to $K-\bar{K}$ mixing in this limit, as expected.

Yet, in an actual scenario, the masses of the $A^0,~H^0$ fields should be set at some reasonable scale. Also, the CP-even mixing angle $\alpha$ deviates somehow from the saturation limit.
To get insight into model constraints from $K-\bar{K}$ mixing, we consider the simplified scenario where $m_{A^0}\gg m_{h^0}$ and $Y^{(2)}_{d12}=0$; in this case, $\widetilde{C}_2=C_4=0$.  As we are close to the decoupling limit, we write $\alpha=\beta-\pi/2+\epsilon$, where $\epsilon\ll 1$, and we may keep only the first term in (\ref{c2}) due to a cancellation between the $H^0$ and $A^0$ contributions.  This approximation is valid up to a $\mathcal{O}(1)$ factor, and should be sufficient for our purposes.
In this limit, the nonvanishing matching conditions become
\be
C_2(m_{h^0})=-\frac{1}{4}(Y^{(2)*}_{d21})^2\left(\frac{\epsilon}{\cos\beta}\right)^2 +\mathcal{O}(\epsilon^3)~, \\
\ee

To get the final answers, we must run down to the hadronic scale to resum QCD logarithms and match operator matrix elements to the expressions with bag factors, as described in \cite{BMZ}, for instance.  Using their equations (14-15), we find:
\bea
C_2(\mu_{\rm had})&=&\eta_{22}C_2(m_{h^0})~, \nn \\
C_3(\mu_{\rm had})&=&\eta_{32}C_2(m_{h^0})~, \nn \\
\eea
and all others zero, where
\bea
\eta_{22}&=&0.983\eta^{-2.42}+0.017\eta^{2.75}~, \nn \\
\eta_{32}&=&-0.064\eta^{-2.42}+0.064\eta^{2.75}~, \nn \\
\eea
and
\be
\eta=\left(\frac{\alpha_s(m_c)}{\alpha_s(\mu_{\rm had})}\right)^{6/27}\cdot\left(\frac{\alpha_s(m_b)}{\alpha_s(m_c)}\right)^{6/25}\cdot\left(\frac{\alpha_s(m_{h^0})}{\alpha_s(m_b)}\right)^{6/23}~.
\ee
We choose $\mu_{\rm had}$ to be where $\alpha_s(\mu_{\rm had})=1$ and defining nonperturbative matrix elements at this scale
\bea
\left.\langle K|\mathcal{O}_2|\bar{K}\rangle\right|_{\mu_{\rm had}}&=&-\frac{5}{24}\left(\frac{m_K}{m_d+m_s}\right)^2m_Kf_K^2B_2~, \nn \\
\left.\langle K|\mathcal{O}_3|\bar{K}\rangle\right|_{\mu_{\rm had}}&=&\frac{1}{24}\left(\frac{m_K}{m_d+m_s}\right)^2m_Kf_K^2B_3~, \nn \\
\eea
we can put constraints on the size of $Y^{(2)*}_{d21}$ and $\epsilon$ given $m_{h^0}$.  Here $B_i$ are the bag factors; in what follows, we set $B_i=1$, the ``vacuum saturation approximation," which is sufficient at this level of accuracy.

For example, we can assume $m_{h^0} = 120$ GeV, as suggested by the EW fits and direct searches, and apply constraints on $\Delta m_K$
\be
\Delta m_K=2{\rm Re}\big(\langle K|\mathcal{L}_{\rm eff}|\bar{K}\rangle\big)<3.48\times 10^{-12}~{\rm MeV}~.\label{dmk}
\ee
For simplicity, we let the Yukawa phases vanish\footnote{The introduction of phases would naively weaken the bounds by allowing for destructive interference, so by setting phases to zero gives us the most conservative bound.}.  To satisfy (\ref{dmk}) we require that $|\epsilon|<10^{-5}$ for $\mathcal{O}(1)$ or slightly smaller values of the off-diagonal Yukawas.

To understand the meaning of this constraint, one can use (\ref{alpha}) and a bit of mathematical analysis to find
\be
\epsilon\sim\sin(4\beta)m_{h^0}^2/m_{A^0}^2~.
\ee
For $\tan\beta=20$ and $m_{h^0}=120$ GeV, this means that the heavy Higgses should have masses around 10 TeV or higher.
Yet, due to condition~(\ref{q61}) $Y^{(2)}_{d21}$ is driven to be significantly less than
one. Then the bound on $\epsilon$ may be about two orders of magnitude weaker
($\epsilon \lesssim 10^{-3}$), or the Heavy Higgses may have masses around 1~TeV.

Of course, these bounds should be taken with an appropriate grain of salt, since we should also include the $1/m_{A^0}^2$ terms in the matching conditions, as well as perform a more careful scan over the full parameter space.  However, this simplified analysis gives us a good place to start, and a more careful analysis is reserved for future work \cite{paper2}.

\section{Discussion and Future Work}\label{Concl}
\renewcommand{\theequation}{5.\arabic{equation}}
\setcounter{equation}{0}

In this paper we have attempted to explain the flavor hierarchy by appealing to the two Higgs doublet model.  We have found that we can explain the fermion masses quite easily with little or no hierarchies in the dimensionless Yukawa couplings so long as our Yukawa matrices satisfy two flavor basis independent conditions
\begin{gather}
{\det}_2(Y^{(2)}Y^{(2)\dagger})=0 ~, \\
|\det(Y)|=|\det(Y^{(1)})|~,
\end{gather}
where $Y$ is given by (\ref{q3}).  With these conditions, the Yukawa couplings need at most a 10\% tuning, as opposed to a tuning of one part in $10^6$ in the usual SM.
Furthermore, we have shown that although this model has tree level flavor changing neutral currents, they are all proportional to $Y^{(2)}$ matrix elements in the mass basis which are naturally small in this setup.  The first condition implies that this matrix has (at least) two vanishing eigenvalues, and this motivated us to define a basis where only the 33 component of this matrix was nonzero, which we call ``basis (b)."  This basis may or may not be related to the gauge basis, which is relevant for deriving the CKM matrix, but the conditions we impose are basis independent and therefore will hold everywhere, including the physical mass basis.

This paper has taken these conditions as axioms of the flavor sector, but it is certainly within the realm of possibility  \cite{FN, FN2, FN3, FN4, FN5, FN6, FN7, FN8} that there is a dynamical explanation for this Yukawa pattern.  For example, one might imagine that the Yukawa matrices are actually vevs of fields that are charged under some larger flavor symmetry which is spontaneously broken at some high scale.  Then this pattern can come from minimizing some as yet unknown effective potential, and technical naturalness of the couplings will protect the pattern as we run to lower scales.  Such possible UV completions will be considered in future work.

Typically the most important flavor changing standard candle is $K-\bar{K}$ mixing due to the high precision of the measurements.  We considered the simple case of the near-decoupling limit in the vacuum saturation approximation, where only the light Higgs boson contributes appreciatively to the mixing parameters.  We estimate that as long as the heavy Higgs states are around  a TeV or higher, there are no significant contributions to this observable.  Since we remain agnostic on what mechanism stabilizes the Higgs masses, we do not view this as a problem from the flavor puzzle point of view.  Generalizing this to other points in Higgs parameter space is straightforward and will be considered in more detail in a followup paper \cite{paper2}.  In addition, it is a straightforward exercise to repeat the analysis for $D-\bar{D}$~\cite{Dmix} and $B-\bar{B}$~\cite{Bmix}  mixing as well.  Each of these are sensitive to different $Y^{(2)}_{ij}$, and together, along with the above condition, can be used to test the full validity of this model.  For the lepton sector, $\mu-e$ conversion, as well as rare $\mu$ and $\tau$ decays can also be used.  We will study these constraints in \cite{paper2}.

One can also imagine solving the larger Higgs fine tuning problem with some extended model such as supersymmetry.  If one wishes to incorporate this model into the MSSM, we would require four Higgs doublets.  Then there would be a basis analogous to our basis (b) where two of these Higgs doublets only coupled to the heavier generations, and the other pair of Higgs doublets coupled to all three generations, where each pair would have an up-type and a down-type Higgs.  It would be interesting to see what analogous constraints we would have to put on the corresponding Yukawa matrix elements in such a model.

Another interesting task would be to test how our model works for the neutrino sector, provided that neutrino masses or
their ratios (rather than mass differences) are known, and all the neutrino mass terms (beyond the
Yukawa sector) are specified.

Finally, there are other phenomenological questions we can ask in this model of the Higgs sector.  For example, the important decay $h\rightarrow\gamma\gamma$ is tyically dominated by top and W/Z particles in a loop.  But with the possibility of changing the Yukawa couplings, this can have strong effects on this decay and possibly change the expectations for discovery at the LHC.  We will discuss this in more detail in \cite{paper2}.

\section*{Acknowledgements}

We would like to thank Yossi Nir for helpful
remarks during the ``Indirect Searches for New Physics at the time of LHC" program
at the Galileo Galilei Institute for Theoretical Physics in Florence (Italy). Support was
provided by the U.S.\ Department of Energy under Contract DE-FG02-96ER41005.
A.A.P was also supported by the U.S.\ National Science Foundation under
CAREER Award PHY--0547794.

\appendix
\section{The Higgs sector} \label{higgs}
\renewcommand{\theequation}{A.\arabic{equation}}
\setcounter{equation}{0}

In this appendix we review the structure of the Higgs sector.  We have two Higgs doublets:
\be
\Phi_i=\cvec{\phi_i^+}{\phi_i^0}\qquad i=1,2~.
\ee
We can write a generic potential for the these fields:
\bea
\mathcal{V}&=&\mu_1^2\Phi^\dagger_1\Phi_1+\mu_2^2\Phi^\dagger_2\Phi_2+\mu_3^2(\Phi_1^\dagger\Phi_2+{\rm h.c.})+\frac{\lambda_1}{2}(\Phi_1^\dagger\Phi_1)^2+\frac{\lambda_2}{2}(\Phi_2^\dagger\Phi_2)^2 \nn \\
& &+~\lambda_3(\Phi_1^\dagger\Phi_1)(\Phi_2^\dagger\Phi_2)+\lambda_4(\Phi_1^\dagger\Phi_2)(\Phi_2^\dagger\Phi_1)+\Big(\frac{\lambda_5}{2}(\Phi_1^\dagger\Phi_2)^2+{\rm h.c.}\Big) \nn \\
& &+~\Big(\lambda_6(\Phi_1^\dagger\Phi_1)(\Phi_1^\dagger\Phi_2)+\lambda_7(\Phi_2^\dagger\Phi_2)(\Phi_1^\dagger\Phi_2)+{\rm h.c.}\Big)~. \label{V}
\eea
One can easily check that the $\lambda_{6,7}$ terms
introduce no essential change  in the analysis \cite{aim1},
thus they may be neglected for simplicity.  We also assume that
$\lambda_5$ and $\mu_3^2$ are real: thus there is no explicit CP-violation in the
Higgs potential. Also, no spontaneous CP-violation is assumed, thus the Higgs doublet
vacuum expectation values are taken to be real.

The Higgs doublet
vacuum states may be presented in the following form:
\begin{equation}
\langle\Phi_1\rangle = \frac{1}{\sqrt{2}} \left(\begin{array}{c}
0 \\ v_1
\end{array} \right),  \hspace{0.5cm}
\langle\Phi_2\rangle = \frac{1}{\sqrt{2}} \left(\begin{array}{c}
0 \\ v_2
\end{array} \right)~, \label{m2a}
\end{equation}
 with $v_{1,2} > 0$ and real. The Higgs potential minimum conditions,
 \begin{equation}
 \frac{\partial \mathcal{V}}{\partial v_1} =  \frac{\partial \mathcal{V}}{\partial v_2} = 0~,
 \label{m3}
 \end{equation}
 may be written as
 \begin{eqnarray}
 \lambda_1 v_1^2 + \tilde{\lambda} v_2^2 + 2 \mu_1^2 +
 2 \mu_3^2 v_2/v_1 = 0~, \label{m4} \\
 \lambda_2 v_2^2 + \tilde{\lambda} v_1^2 + 2 \mu_2^2 +
 2 \mu_3^2 v_1/v_2 = 0~, \label{m5}
 \end{eqnarray}
where $\tilde{\lambda} = \lambda_3 + \lambda_4 + \lambda_5$.

The Higgs doublet vacuum expectation values must satisfy the
following condition: $v_1^2 + v_2^2 = v^2 = (246~{\rm GeV})^2$.  Constraints on
the coupling  constants $\lambda_i$ may be derived from the analysis
of their renormalization group equations \cite{aim1, aim2}.  Two of the
mass parameters of the Higgs potential, say $\mu_1^2$ and $\mu_2^2$, may be
eliminated from minimum conditions (\ref{m4}) and (\ref{m5}). The parameter
$\mu_3^2$ however remains arbitrary.

It should be mentioned at this point that in a general Type-III two-Higgs doublet model, $v_1$ and $v_2$ are not well defined \cite{DH,HO}.  In fact, since $\Phi_{1,2}$ have the same quantum numbers, any linear combination of them can get a vev, and one can always perform a field redefinition that changes the value of $v_{1,2}$ while keeping the value of $v^2=v_1^2+v_2^2$ fixed.  However, when we discuss Higgs couplings to the fermions in Sections \ref{TwoGen} and \ref{ThreeGen}, in particular conditions (\ref{q6}), (\ref{q42}) and (\ref{q61}), this ambiguity is removed, and so we will proceed as if these vevs have a physical meaning.

One may express $\Phi_1$ and $\Phi_2$ in terms of the excited Higgs states in the
following from:
\begin{equation}
\Phi_1 = \left( \begin{array}{c}
 G^+ \cos{\beta} - H^+ \sin{\beta} \\
\frac{1}{\sqrt{2}} \left[v_1  + h_1 + i \left( G^0 \cos{\beta} - A^0 \sin{\beta}
 \right) \right] \end{array} \right)~, \label{m6}
\end{equation}
\begin{equation}
\Phi_2 = \left( \begin{array}{c}
 G^+ \sin{\beta} + H^+ \cos{\beta} \\
\frac{1}{\sqrt{2}} \left[v_2  + h_2 + i \left(G^0 \sin{\beta +
A^0 \cos{\beta}} \right) \right] \end{array} \right)~,
\label{m7}
\end{equation}
where $\tan{\beta} = v_2/v_1$, $G^0$, $G^{\pm}$ are the Goldstone modes,
$h_1$, $h_2$ are CP-even, $A^0$ is CP-odd and $H^{\pm}$ is the charged
physical Higgs states.  It is straightforward to check, using minimum conditions (\ref{m4})
and (\ref{m5}),
that the Higgs potential contains no
terms linear in the physical Higgs fields.

Without any CP violation, the CP-even and -odd Higgs states will not mix, and can be considered separately.  The mass of the CP-odd Higgs boson is given by
\begin{equation}
m_{A^0}^2 = \frac{- 2 \mu_3^2}{\sin{2 \beta}} - \lambda_5 v_2^2 =
\mu_1^2 + \mu_2^2 + \frac{1}{2} \left( \lambda_1
\cos^2{\beta} + \lambda_2 \sin^2{\beta} + \lambda^\prime\right) v^2~,
\label{m8}
\end{equation}
where $\lambda^\prime = \lambda_3 + \lambda_4 - \lambda_5$. The CP-odd mass
may be chosen to be a free parameter of the theory.
Then the charged Higgs mass is given by
\begin{equation}
m_{H^\pm}^2 = m_{A^0}^2 - \frac{(\lambda_4 - \lambda_5) v^2}{2}~. \label{m9}
\end{equation}

The $2 \times 2$ mass matrix for the CP-even Higgs fields $h_1$ and $h_2$ is
the following:
\bea
M^2 &=& \left( \begin{array}{cc}
 \left(\lambda_1 \cos^2{\beta} + \lambda_5 \sin^2{\beta} \right) v^2 +
m_{A^0}^2 \sin^2{\beta} &
\left((\lambda_3 + \lambda_4) v^2 - m_{A^0}^2 \right) \sin{\beta} \cos{\beta} \\
\left((\lambda_3 + \lambda_4) v^2 - m_{A^0}^2 \right) \sin{\beta} \cos{\beta} &
\left(\lambda_2 \sin^2{\beta} + \lambda_5 \cos^2{\beta} \right) v^2 +
m_{A^0}^2 \cos^2{\beta} \end{array} \right)~. \nn\\
\label{m10}
\eea
The CP-even Higgs eigenstates, $h^0$, $H^0$, are related to $h_1$ and $h_2$ as
\begin{eqnarray}
&& H^0 = h_1 \cos{\alpha} + h_2 \sin{\alpha}~, \label{m11} \\
&& h^0 = - h_1 \sin{\alpha} + h_2 \cos{\alpha}~,   \label{m12}
\end{eqnarray}
where
\be
\tan{2\alpha}=\frac{2 M_{12}^2}{M_{11}^2 - M_{22}^2}~, \label{alpha}
\ee
and
\begin{equation}
m_{h^0,H^0}^2 = \frac{1}{2} \left[ M_{11}^2 + M_{22}^2 \mp
\sqrt{(M_{11}^2 - M_{22}^2)^2
+ 4 (M_{12}^2)^2} \right]~. \label{m15}
\end{equation}

Writing explicitly the matrix elements in (\ref{alpha})-(\ref{m15})
would make these formulae rather complicated -- due to large number of independent
couplings the predictive power of the general two-Higgs doublet model is rather weak. Nevertheless,
one can derive an upper bound on the lightest CP-even Higgs mass
\begin{equation}
m_{h^0}^2 \leq \left(\lambda_1 \cos^4{\beta} + \lambda_2 \sin^4{\beta} +
2 \tilde{\lambda} \sin^2{\beta} \cos^2{\beta} \right) v^2~, \label{m16}
\end{equation}
which is saturated as $m_{A^0}^2 \to \infty$; this state is usually identified with the ``Standard Model Higgs."  In the same limit, $m_{H^0}^2
\approx m_{H^\pm}^2 \approx m_{A^0}^2$, that is to say all the other Higgs particles
may be arbitrarily heavy. Also, at this limit the mixing angle is given by
$\alpha \approx \beta - \pi/2$.

Note that for $m_{A^0}^2 \gg v^2$, the problem of flavor changing neutral currents
is avoided in a natural way. The FCNCs are suppressed when $A^0$ or $H^0$ is exchanged.
One can also show  that for $\alpha = \beta - \pi/2$, no FCNCs occur when quarks interact with
the exchange of the lightest Higgs boson $h^0$.  This result is intuitive, since in this limit we effectively only have one Higgs doublet as in the usual SM, and there are no FCNCs coming from the SM Higgs sector.

\section{$Y$ and $Y Y^\dagger$ in basis~(b)}\label{details1}
\renewcommand{\theequation}{B.\arabic{equation}}
\setcounter{equation}{0}

For the three generation case, in basis~(b) the total Yukawa matrix is given
by
\begin{equation}
Y^b = \left(\begin{array}{ccc}
Y^{(1)}_{11} \  \ & Y^{(1)}_{12} \ \ & Y^{(1)}_{13} \\
Y^{(1)}_{21} \  \ & Y^{(1)}_{22} \ \ & Y^{(1)}_{23} \\
Y^{(1)}_{31} \  \ & Y^{(1)}_{32} \ \ & ~Y^{(1)}_{33} + y_3^{(2)} \tan{\beta}
\end{array} \right) \label{q64}
\end{equation}
The elements of the Hermittean matrix $Y Y^\dagger$ in the same basis are
\begin{eqnarray}
\nonumber
&& \left(Y^b Y^{b \dagger} \right)_{11} =
|Y^{(1) b}_{11}|^2 + |Y^{(1) b}_{12}|^2 + |Y^{(1) b}_{13}|^2 \\
\nonumber
&& \left(Y^b Y^{b \dagger} \right)_{21} =
\left(Y^b Y^{b \dagger} \right)_{12}^\star =
Y^{(1) b}_{21} Y^{(1) b \star}_{11} +
Y^{(1) b}_{22} Y^{(1) b \star}_{12} +
Y^{(1) b}_{23} Y^{(1) b \star}_{13} \\
\nonumber
&& \left(Y^b Y^{b \dagger} \right)_{31} =
\left(Y^b Y^{b \dagger} \right)_{13}^\star =
Y^{(1) b}_{31} Y^{(1) b \star}_{11} +
Y^{(1) b}_{32} Y^{(1) b \star}_{12} +
\left(Y^{(1) b}_{33} + y_3^{(2)} \tan{\beta}
\right) Y^{(1) b \star}_{13}
\\
&& \left(Y^b Y^{b \dagger} \right)_{22} =
|Y^{(1) b}_{21}|^2 + |Y^{(1) b}_{22}|^2 +
|Y^{(1) b}_{23}|^2  \label{q65} \\
\nonumber
&& \left(Y^b Y^{b \dagger} \right)_{32} =
\left(Y^b Y^{b \dagger} \right)_{23}^\star =
Y^{(1) b}_{31} Y^{(1) b \star}_{21} +
Y^{(1) b}_{32} Y^{(1) b \star}_{22} +
\left(Y^{(1) b}_{33} + y_3^{(2)}
\tan{\beta} \right) Y^{(1) b \star}_{23} \\
\nonumber
&& \left(Y^b Y^{b \dagger} \right)_{33} =
|Y^{(1) b}_{31}|^2 + |Y^{(1) b}_{32}|^2 + |Y^{(1) b}_{33}|^2 +
2 Re\left[ y_3^{(2)} Y^{(1) b \star}_{33} \right] \tan{\beta} +
|y_3^{(2)}|^2 \tan^2{\beta}
\end{eqnarray}

\section{Mass eigenvalues and ratios in terms of Isospin basis Yukawa couplings}\label{details2}
\renewcommand{\theequation}{C.\arabic{equation}}
\setcounter{equation}{0}

The mass matrix eigenvalues can be written in terms of the Yukawa couplings.  To leading order in $\tan\beta$ the results are:

{\footnotesize
\be
|y_3|^2 \approx
\Tr\left( Y^{(2)} Y^{(2) \dagger} \right) \tan^2{\beta} \label{q75}
\ee
\begin{align}
|y_2|^2 \approx& \nn \\
&\frac{{\det}_2\left(Y^{(1)} Y^{(2) \dagger} +
Y^{(2)} Y^{(1) \dagger} \right) + \sum_{i \neq j}\left[\left( Y^{(1)}
Y^{(1) \dagger} \right)_{ii} \left( Y^{(2)} Y^{(2) \dagger} \right)_{jj} -
\left( Y^{(1)}
Y^{(1) \dagger} \right)_{ij} \left( Y^{(2)} Y^{(2) \dagger} \right)_{ji}\right]}
{\Tr\left( Y^{(2)} Y^{(2) \dagger} \right) } \label{q76} \nn\\
\end{align}
\begin{align}
|y_1|^2 \approx&\nn \\
& \frac{\det\left( Y^{(1)} Y^{(1) \dagger} \right)
\tan^{-2}{\beta}}
{{\det}_2\left(Y^{(1)} Y^{(2) \dagger} +
Y^{(2)} Y^{(1) \dagger} \right) + \sum_{i \neq j}\left[\left( Y^{(1)}
Y^{(1) \dagger} \right)_{ii} \left( Y^{(2)} Y^{(2) \dagger} \right)_{jj}  -
\left( Y^{(1)}
Y^{(1) \dagger} \right)_{ij} \left( Y^{(2)} Y^{(2) \dagger} \right)_{ji}\right]} \label{q77}
\end{align}}
\vspace{1cm}
Subsequently, for the mass ratios, $m_{q_i}/m_{q_j} = |y_i|/|y_j|$, one gets
{\footnotesize
\begin{align}
\frac{m_{q_2}}{m_{q_1}} \approx&\frac{\tan\beta}{\sqrt{\Tr\left( Y^{(2)} Y^{(2) \dagger} \right) \ \det\left(Y^{(1)} Y^{(1) \dagger} \right)}}~\times \nn \\
&\left[ {\det}_2\left(Y^{(1)} Y^{(2) \dagger} +
Y^{(2)} Y^{(1) \dagger} \right) + \sum_{i \neq j}\left[\left( Y^{(1)}
Y^{(1) \dagger} \right)_{ii} \left( Y^{(2)} Y^{(2) \dagger} \right)_{jj} -
\left( Y^{(1)}
Y^{(1) \dagger} \right)_{ij} \left( Y^{(2)} Y^{(2) \dagger} \right)_{ji}\right] \right]
\label{q78}
\end{align}
\begin{align}
\frac{m_{q_3}}{m_{q_2}} \approx&\nn\\
&\frac{\Tr\left( Y^{(2)} Y^{(2) \dagger} \right) \tan{\beta}}
{\sqrt{{\det}_2\left(Y^{(1)} Y^{(2) \dagger} +
Y^{(2)} Y^{(1) \dagger} \right) + \sum_{i \neq j}\left[\left( Y^{(1)}
Y^{(1) \dagger} \right)_{ii} \left( Y^{(2)} Y^{(2) \dagger} \right)_{jj} -
\left( Y^{(1)}
Y^{(1) \dagger} \right)_{ij} \left( Y^{(2)} Y^{(2) \dagger} \right)_{ji}\right]}} \label{q79}
\end{align}
\begin{align}
\frac{m_{q_3}}{m_{q_1}} \approx &\sqrt{\frac{\Tr\left( Y^{(2)} Y^{(2) \dagger} \right)}
{\det\left(Y^{(1)} Y^{(1) \dagger} \right)}} \tan^2{\beta}~\times\nn \\
&\sqrt{{\det}_2\left(Y^{(1)} Y^{(2) \dagger} +
Y^{(2)} Y^{(1) \dagger} \right) + \sum_{i \neq j}\left[\left( Y^{(1)}
Y^{(1) \dagger} \right)_{ii} \left( Y^{(2)} Y^{(2) \dagger} \right)_{jj} -
\left( Y^{(1)}
Y^{(1) \dagger} \right)_{ij} \left( Y^{(2)} Y^{(2) \dagger} \right)_{ji}\right]}
 \label{q80}
\end{align}}


%
\end{document}